\documentclass[sigconf]{acmart}
\AtBeginDocument{%
  \providecommand\BibTeX{{%
    \normalfont B\kern-0.5em{\scshape i\kern-0.25em b}\kern-0.8em\TeX}}}

\acmConference[IUI'23]{IUI'23: 28th International Conference on Intelligent User Interfaces}{March 27 -- 31, 2023}{Sydney, Australia}
\acmBooktitle{IUI'23: 28th International Conference on Intelligent User Interfaces,
  March 27 -- 31, 2023, Sydney, Australia}

\copyrightyear{2023}
\acmYear{2023}
\setcopyright{rightsretained}
\acmConference[IUI '23 Companion]{28th International Conference on Intelligent
User Interfaces}{March 27--31, 2023}{Sydney, NSW, Australia}
\acmBooktitle{28th International Conference on Intelligent User Interfaces (IUI
'23 Companion), March 27--31, 2023, Sydney, NSW, Australia}
\acmDOI{10.1145/3581754.3584111}
\acmISBN{979-8-4007-0107-8/23/03}

\begin{document}

\title[Adapting Large Language Model and Gamification in ITS to Scaffold CS1 Courses]{Leveraging Large Language Model and Story-Based Gamification in Intelligent Tutoring System to Scaffold Introductory Programming Courses: A Design-Based Research Study}



\author{Chen Cao}
\email{ccao5@sheffield.ac.uk}
\orcid{0000-0003-4368-0336}
\affiliation{%
  \institution{University of Sheffield}
  \country{United Kingdom}
  \postcode{S10 2TN}
}




\begin{abstract}
Programming skills are rapidly becoming essential for many educational paths and career opportunities. Yet, for many international students, the traditional approach to teaching introductory programming courses can be a significant challenge due to the complexities of the language, the lack of prior programming knowledge, and the language and cultural barriers. This study explores how large language models and gamification can scaffold coding learning and increase Chinese students’ sense of belonging in introductory programming courses. In this project, a gamification intelligent tutoring system was developed to adapt to Chinese international students’ learning needs and provides scaffolding to support their success in introductory computer programming courses. 

My research includes three studies: a formative study, a user study of an initial prototype, and a computer simulation study with a user study in progress. Both qualitative and quantitative data were collected through surveys, observations, focus group discussions and computer simulation. The preliminary findings suggest that GPT-3-enhanced gamification has great potential in scaffolding introductory programming learning by providing adaptive and personalised feedback, increasing students’ sense of belonging, and reducing their anxiety about learning programming.
\end{abstract}


\newcommand{\todo}[1]{{\color{blue} #1}}

\maketitle

\section{Problem Statement}

Computing and technology are increasingly ubiquitous and have become a necessary part of many educational paths, professional opportunities, and industries \cite{pedro2019artificial}. Consequently, the importance of programming knowledge and coding skills has grown significantly in recent years. Introductory programming courses, often referred to as computer science 1 (CS1) courses are designed to introduce students to the fundamentals of programming and coding \cite{becker201950}. However, these courses can be challenging for many international students, who may lack prior programming expertise and are confronted with a language and cultural barriers\cite{khanal2019challenges}.

Recent years have also seen an increased number of Chinese international students enrolling in universities in the UK\cite{Wit2020InternationalizationOH}. Despite the clear benefits of international student enrollment, Chinese international students can often feel isolated in their new environment and struggle to integrate into university life\cite{chen2019chinese}. In the context of programming education, Chinese international students often face hardship due to a lack of prior exposure to coding concepts and language \cite{Alaofi2022}. As a result, they are more likely to experience anxiety and a feeling of low belonging. However, previous research showed bias and stereotypes when describing Chinese students’ learning behaviours in global higher education \cite{Heng2018}. Chinese international students are often perceived as passive and reluctant learners when adapting to British educational systems \cite{zhu2022shhhh}. This stereotype, however, fails to take into account the cultural and educational factors driving these behaviours. 

At the same time, gamification has seen a surge in popularity in the educational sector. It has been used as a tool to engage students and encourage learning in a range of contexts \cite{welbers2019gamification}. This presents an opportunity to investigate the potential of gamification to scaffold coding learning and increase the sense of belonging among Chinese international students in introductory programming courses. Additionally, the use of educational technologies, such as intelligent tutoring systems in educational settings has been growing steadily in recent years, as an increasing number of educators recognize the potential of this approach to engage and motivate learners \cite{szymkowiak2021information}. An intelligent tutoring system is particularly attractive due to its ability to personalise the learning experience, offering tailored activities to meet the needs of diverse learners \cite{shemshack2021comprehensive}. In addition, an intelligent tutoring system can provide feedback in a timely and effective manner, offering learners the opportunity to improve their skills in a supportive and engaging environment \cite{alshaikh2021ai}. 

The research background of this thesis is rooted in the idea that AI-enhanced gamification can be used to scaffold learning and increase belonging among Chinese international students in introductory programming courses. AI-enhanced gamification combines the use of gamification techniques with AI-driven intelligent tutoring systems to create a personalised learning experience. This approach encourages learners to engage in the learning process and offers them the opportunity to practise and improve their coding skills in a supportive and engaging environment.

This research is motivated by the need to bridge the gaps in programming knowledge between Chinese international students and their domestic peers. It is also driven by the urgent need to create an inclusive learning environment that is accessible and welcoming to all students. The research findings could have a considerable impact on the teaching and learning of programming for Chinese international students. It has the potential to provide insight into how AI-enhanced gamification can be used to make the learning process easier and more effective for these students, as well as to foster a sense of belonging. The findings could inform instructors, administrators, and policymakers of the most effective strategies for teaching and learning to program for this population. This study can also be expanded to support other underrepresented student groups (such as female STEM students and other international students from different countries and cohorts) who also need a sense of belonging to their peers, faculty, and subject-related careers.

\section{Research aim and questions}

The main focus of this research project is to explore the potential of AI-enhanced gamification to scaffold coding learning and increase the sense of belonging among Chinese international students in introductory programming courses. Specifically, the focus will be on designing, evaluating, and refining the use of AI-enhanced gamification to improve learning outcomes and increase motivation among Chinese international students in introductory programming courses.

The research project is framed within the context of design-based research, which emphasises the importance of designing, implementing, and evaluating a learning environment, with the goal of optimising the learning experience. The research project is guided by two research questions: 

\begin{itemize}
    \item RQ1: what are the challenges that Chinese international students are facing regarding developing a sense of belonging and code learning in introductory programming courses?

     \item RQ2: how can AI and gamification be used to scaffold coding learning and increase the sense of belonging among Chinese international students in introductory programming courses?
\end{itemize}

\section{Related work}

This research is situated within the larger context of educational technology and game-based learning. The use of technology in education has become increasingly popular, with a growing number of educators recognising the potential of this approach to engage and motivate learners \cite{szymkowiak2021information}. This study specifically focuses on the application of AI-enhanced gamification to scaffold coding learning among Chinese international students. 

There has been a considerable amount of research examining the challenges and barriers associated with teaching introductory programming courses, e.g. \cite{alam2022platform}.
These studies have highlighted the need for more effective pedagogical approaches to engage learners in CS1 courses. Several researchers have proposed the use of game-based learning techniques such as serious games and simulations to increase engagement and motivation among students \cite{papadakis2019evaluating}
. Gamification combines elements of gaming (e.g., points, rewards, leaderboards) with traditional educational activities to create an engaging learning experience. 

The use of intelligent tutoring systems (ITS) in programming education has also been studied extensively. Research has shown that ITS can improve student performance and reduce cognitive load in programming courses. For example, Grenander et al. \cite{Grenander2021} proposed an AI-enhanced educational system that was designed to provide personalised feedback based on individual learners’ needs and evaluated its effectiveness using deep discourse analysis. Similarly, Eguchi \cite{eguchi2021contextualizing} investigated the use of AI-enhanced games to support STEM education for children with visual impairments. They found that the AI-enhanced game had a positive impact on engagement and motivation among participants. Furthermore, ITS can provide personalised instruction and targeted remediation, which can be particularly beneficial for those who lack prior experience in programming. In particular, GPT-3 (Generative Pre-trained Transformer 3), an advanced language model developed by OpenAI, has been used to improve the performance of ITSs by providing natural language understanding capabilities\cite{Tack2022TheAT}. 

This research project is also informed by recent studies on belonging in higher education. It has been argued that belonging is an important factor when it comes to understanding and supporting the learning experiences of international students \cite{Cureton2019WeBD}. Studies have also shown that international students often face difficulties in developing a sense of belonging due to language and cultural barriers \cite{Cena2021SenseOB}. 

\section{Research method}

This research project applies design-based research (DBR) as its methodological approach. DBR is an iterative process that
emphasises the importance of designing, implementing, and evaluating educational technology to optimise the learning experience. This approach has been widely used in educational settings to investigate the effectiveness of new technologies and approaches in engaging and motivating learners. The use of DBR provides
a unique opportunity to combine theoretical insights with practical implementation to create meaningful interventions that are responsive
to the needs of diverse learners.

The research is divided into three phases: 1) a survey to identify Chinese students' needs and the challenges they met regarding a sense of belonging and learning experience in introductory programming courses; 2) a design probe with a story-based gamification prototype to increase Chinese students’ sense of belonging and improve their learning experience in CS1 courses with user evaluation; 3) the development and systematic evaluations of an intelligent tutoring system leveraging large language model and gamification features for CS1 courses.

The study used a mixed-methods approach, incorporating both quantitative and qualitative data. The quantitative data were collected through a survey measuring students’ sense of belonging, academic performance, and academic emotions, and a computer simulation study evaluating the performances of the ITS. The qualitative data were collected through participatory observations and focus group interviews to explore students' experiences and perceptions in more depth. The data were analyzed using descriptive statistics, thematic analysis, and computational analysis. 

\section{Preliminary results}
\subsection{Study 1: Formative study}
The initial survey was conducted as a formative study in the first semester of the 21/22 academic year. Based on the questionnaire with 57 Chinese international students and in-depth interviews with nine of them, the study found a number of unique challenges faced by the participants in the CS1 courses. Chinese students generally found it difficult to cope with the demands of the programming course. A number of factors were identified as contributing to this difficulty, including the challenges to adapt to new teaching methods emphasizing independent thinking, critical thinking, and innovative learning; less interaction with teachers and classmates; disconnection between the knowledge and real-life cases and not receiving enough academic support with timely help and real-time feedback. As a result, they have  low engagement in classes, low retention and negative academic emotions. One of the key factors that have been identified as contributing to these unsmooth academic transition experiences is a lack of intrinsic motivation, especially a lack of sense of belonging. 

\subsection{Study 2: A prototype of story-based ITS}
The second study designed, deployed and evaluated a prototype of a story-based gamification design on a learning management system Blackboard. Based on the initial findings from the survey, the prototype adopted a set of gamification features, including animated trailers, story-telling, role-play, teamwork, points, leaderboards and feedback. The study was conducted over a period of two weeks with a total of 34 Chinese students enrolled in the introductory programming course INF6032 Big Data Analytics, which is one of the core modules of the MSc in Data Science programme at the University of Sheffield. The practical session in week 6 was set in the context of airline companies’ survival during the pandemic. In week 7, the story was about solving a criminal case. The narrative storytelling of each session with real-life cases was designed to make students feel more connected to the programme.

The findings from questionnaires (n=32), focus groups (n=32) and participatory observation revealed that Chinese students generally had positive perceptions of the effect of story-based gamification design on their sense of belonging in the introductory programming course. The majority of students reported feeling more motivated to learn, more engaged in the course, and more connected to their classmates as a result of the story-based gamification design. In addition, the findings suggested that story-based gamification design had a positive effect on Chinese students’ sense of belonging in the introductory programming course. Most students reported feeling a greater sense of belonging in the course after the story-based gamification design was implemented. It was also found that Chinese students have some unique cultural needs which should be considered in the story-based gamification design. These findings suggest that story-based gamification design can be an effective means of improving Chinese students’ sense of belonging in the introductory programming course.

\subsection{Study 3: A model of the story-based ITS empowered by GPT-3}

The reflection upon the findings from the previous study encouraged the development of the current model of the story-based intelligent tutoring system (ITS). The ITS empowered by GPT-3 was designed to better fit the needs of individual learners, and increase their sense of belonging to the class, institution and career-related subject, in order to further foster inclusion in higher education.  

The design of the user interface is inspired by the English TV series Sherlock with abundant modern English elements, which is ideal to engage international students coming to the UK for higher education. The user interface of the ITS was built on a popular open-source template\footnote{Bootstrap 4 UI kit, https://demos.creative-tim.com/now-ui-kit-react/\#/index?ref=nukr-github-readme} in React, a mainstream JavaScript library for developing front-ends. The overall design principles of the web app are inspired by the famous detective, Sherlock Holmes. The colour scheme, typography and navigation are all based on the idea of the Mind Palace, with a focus on the dark blue and white colours to convey a feeling of sophistication and intelligence. 

The development of the learning system consists of four main components: the instructional content, the gamification mechanism design, the user interface, and the generative language model. The learning content consists of the introduction of programming knowledge, demo explanations and exercises to provide the learning materials to the users. The game engine was designed to provide a fun and engaging learning experience. It consisted of a set of gamification elements, including alternative reality, points, badges, personalized feedback and encouragements, levels and challenges, an avatar, a progress bar and an exploratory word cloud. The user interface is responsible for displaying the learning content and gamification mechanisms to the users and accepting user input. The generative language model is used for providing AI capabilities such as Q\&A and explaining code to the system. 

The system is designed to be a web-based application. It consists of three main components, which are the front end, the back end and the GPT-3 platform. The front end used React library to develop the user interface and enable interactions. The back end is responsible for data storage, processing and retrieval. In this system, Firebase is applied to connect with the interface and synchronise data. All the states (user behaviours and inputs) in the front end were centralised with Redux and synchronised to Firebase, where all the user data, such as the user’s avatar, learning progress and chat history with the AI conversational agents were stored for further analysis. 

GPT-3 was used in this system to train the AI model for the intelligent agents. The agents were designed to interact with the users, and provide guidance and support throughout the learning process. In addition, the agents were also responsible for marking the user’s progress and giving feedback. There were four chatbots providing support to students with different functions, including the intelligent tutoring bot \textit{Sherlock}, the peer and critical thinking bot \textit{Watson}, the career engagement bot \textit{Inspector Lestrade} and the emotion support bot \textit{Mrs Hudson}. When the student asked a question, \textit{Sherlock} answered it first. Then \textit{Watson}, \textit{Inspector Lestrade} and \textit{Mrs Hudson} generated follow-up questions to continue the conversation, in order to foster inquiry-based learning and improve students' computational thinking and understanding of job market needs. 

In the interface of marking and feedback, students can submit the codes for quizzes and get real-time and tailored feedback about their answers. The model first extracts the code snippet from the user input and then generates an explanation for the code snippet by calling the GPT-3 API with the code snippet and a prompt as input. The prompt is a natural language question that can be used to generate an explanation for the code snippet. The code-explaining module uses a set of predefined prompts to generate explanations for the code snippets. The prompts are selected to cover a wide range of programming concepts. 

A computer simulation study and a pilot user study were conducted to evaluate the ITS. By using prompt programming, a dataset containing 360 rounds of simulated questions that students may ask in the class and answers were simulated based on the learning objectives and instructors' reflections. The findings from semantic similarity analysis, topic modelling and sentiment analysis indicated that GPT-3 empowered agents performed well in providing feedback and having insightful conversations, but the quality of answers depend on the form and level of the questions. The ITS was also piloted with different stakeholders, including potential users, instructors, practitioners and researchers to assess the usability, acceptability and effectiveness of the system. Findings from the pilot study indicated that the system performs well in terms of usability and acceptability. Most participants reported feeling positive about their experience with the intelligent tutoring system, with many saying they were impressed and felt supported by the AI agents. They also commented positively on the gamification elements in the system, such as the intriguing storyline of Sherlock.

\section{Implications and future work}
The current studies suggested that GPT-3 as a large language model trained on a vast amount of text data is particularly well-suited to answering questions at the remembering and understanding levels of Bloom's taxonomy. These levels involve recalling and comprehending information, which is a strength of large language models like GPT-3. IT also performed well in answering questions at the higher levels of the taxonomy, such as analysis, synthesis, and evaluation if the questions are well-formed. Generally, GPT-3's ability to answer questions in the educational scenario of CS1 courses is promising but limited by its lack of domain-specific knowledge and expertise. 

Despite the promising results of this study, there are still a number of areas for further research and development. Firstly, more detailed studies should be conducted to investigate the long-term impact of AI-enhanced gamification on students’ learning outcomes and sense of belonging in introductory programming courses. Secondly, contextual factors such as culture, language and prior knowledge should be taken into account when designing AI-enhanced gamification systems to better support Chinese international students. Thirdly, more sophisticated machine learning models can be explored to improve the performance of AI agents. Finally, more user studies need to be conducted to explore how AI-enhanced gamification can be used to foster collaboration among team members, reduce anxiety and facilitate learning transfer.

In order to fully realize the potential of the GPT-3 enhanced ITS to increase student sense of belonging in higher education, more research must be conducted in real educational scenarios. More user studies are needed to assess how GPT-3 affects student academic performance, motivation, and sense of belonging. In addition, researchers must continue to refine GPT-3’s capabilities to improve the ITS’s ability to pick up on subtle social cues and to develop a better understanding of student emotions and motivations.

\bibliographystyle{ACM-Reference-Format}
\bibliography{sample-base}

\end{document}